\documentclass{article}
 
\usepackage{amsmath,amssymb,amsfonts}   
\usepackage{graphicx,bm}

\usepackage{float}
\usepackage{epsfig}
\usepackage{esint}

\usepackage{bm,braket}

\newcommand{\calU}{{\mathcal U}}
\newcommand{\calG}{{\mathcal G}}
\newcommand{\calP}{{\mathcal P}}
\newcommand{\calQ}{{\mathcal Q}}
\newcommand{\R}{{\mathbb R}}

\newcommand{\X}{\mathbf{X}}
\renewcommand{\P}{\mathbb{P}}
\newcommand{\PP}{\widetilde{P}}
\newcommand{\x}{\mathbf{x}}

\newcommand{\e}{{\mathrm e}}

\newcommand{\n}{\mathbf n}

\renewcommand{\P}{\mathbb P}
\newcommand{\p}{\widetilde{p}}

\newcommand{\Q}{\widetilde{Q}}
\newcommand{\ellh}{\hat{\ell}}

\begin{document}

 \title{Threshold surface reactions and local time resetting}

\author{\em 
Paul C. Bressloff \\ Department of Mathematics, University of Utah \\
155 South 1400 East, Salt Lake City, UT 84112}
\maketitle

\begin{abstract} 
 In this paper we consider a threshold surface absorption mechanism for a particle diffusing
in a domain containing a single target $\calU $. The target boundary $\partial \calU$ is taken to be a reactive surface that modifies an internal state $Z_t$ of the particle when in contact with the surface at time $t$, with $Z_0=h$. The state $Z_t$ is taken to be a monotonically decreasing function of the so-called boundary local time, and absorption occurs as soon as $Z_t$ reaches zero. (The boundary local time is a Brownian functional that determines the amount of time that the particle spends in a neighborhood of $\partial \calU$.) We first show how to analyze threshold surface absorption in terms of the joint probability density or generalized propagator $P_0(\x,\ell,t|\x_0)$ for the pair $(\X_t,\ell_t)$ in the case of a perfectly reflecting surface, where $\X_t$ and $\ell_t$ denote the particle position and local time at time $t$, respectively, and $\x_0$ is the initial position. 
We then introduce a generalized stochastic resetting protocol in which both the position $\X_t$ and the internal state $Z_t$ are reset to their initial values, $\X_t\rightarrow \x_0$ and $Z_t\rightarrow h$, at a Poisson rate $r$. The latter is mathematically equivalent to resetting the boundary local time, $\ell_t\rightarrow 0$. Since resetting is governed by a renewal process, the survival probability with resetting can be expressed in terms of the survival probability without resetting, which means that the statistics of absorption can be determined by calculating the Laplace transform of $P_0(\x,\ell,t|\x_0)$ with respect to $t$. We contrast this with the case where only particle position is reset, which is not governed by a renewal process. We illustrate the theory using the simple examples of diffusion on the half-line and a spherical target in a spherical domain.

\end{abstract}
%%%%%%%%%%%%%%%%%%%%%%%%%%%

\section{Introduction}

The single-particle version of the diffusion equation represents the evolution of the probability density $p(\x,t|\x_0)$ for the random position $\X_t \in \R^d$ of a Brownian particle at time $t$, given that it started at $\x_0$. Individual trajectories of the particle are generated by a corresponding stochastic differential equation (SDE), which in the case of pure diffusion is given by a Wiener process. It is straightforward to extend the diffusion equation to a bounded domain using classical boundary conditions, namely, Dirichlet ($p=0$), Neumann ($\nabla p=0$) or Robin ($D\nabla p\cdot \n+\kappa_0 p=0$), where $D$ is the diffusivity, $\kappa_0$ is the reactivity constant, and $\n$ is the unit normal at a point on the boundary. However, incorporating such boundary conditions into the associated SDE is more involved. In the case of a totally absorbing boundary (Dirichlet), one simply halts the Brownian motion on the first encounter between particle and boundary. The random time at which this event occurs is known as the first passage time (FPT). On the other hand, a totally or partially reflecting boundary requires a modification of the stochastic process itself. For example, a Neumann boundary condition can be implemented using a Brownian functional known as the boundary local time \cite{Levy39,McKean75,Freidlin85,Majumdar05}, which determines the amount of time that a Brownian particle spends in the neighborhood of a point on the boundary. It is also possible to formulate probabilistic versions of the Robin boundary condition \cite{Papanicolaou90,Milshtein95,Singer08}.

Recently, a theoretical framework for analyzing a more general class of diffusion-mediated surface reactions has been introduced using an encounter-based approach \cite{Grebenkov19b,Grebenkov20,Grebenkov21}. The first step is to consider the joint probability density or propagator $P_0(\x,\ell,t|\x_0)$ for the pair $(\X_t,\ell_t)$ in the case of a perfectly reflecting boundary, where $\X_t$ and $\ell_t$ denote the particle position and local time, respectively. The propagator satisfies a corresponding boundary value problem (BVP), which can be derived using integral representations \cite{Grebenkov20} or path-integrals \cite{Bressloff22a}. The effects of surface reactions are then incorporated 
 by introducing the stopping time 
$
{\mathcal T}=\inf\{t>0:\ \ell_t >\widehat{\ell}\}$,
 with $\widehat{\ell}$ a so-called stopping local time \cite{Grebenkov06,Grebenkov07,Grebenkov20}. Given the probability distribution $\Psi(\ell) = \P[\ellh>\ell]$, the marginal probability density for particle position is defined according to
 $  p_0(\x,t|\x_0)=\int_0^{\infty} \Psi(\ell)P_0(\x,\ell,t|\x_0)d\ell$. It can be shown that the classical Robin boundary condition for the diffusion equation corresponds to the exponential distribution
$\Psi(\ell) =\e^{-\gamma \ell}$, where $\gamma =\kappa_0/D$. One natural generalization of the exponential distribution is obtained by taking the reactivity to depend on the local time $\ell$ (or the number of surface encounters), that is, $\kappa=\kappa(\ell)$. Since one can no longer write down an explicit boundary condition for $p_0(\x,t|\x_0)$, it is necessary to work directly with the propagator.

We recently extended the theory of diffusion-mediated surface reactions to include the effects of stochastic resetting, where the position of the particle is reset to its initial location at a constant resetting rate $r$ \cite{Bressloff22b}. Stochastic resetting was originally studied within the context of diffusion in $\R^d$ and in bounded domains with totally absorbing surfaces \cite{Evans11a,Evans11b,Evans14}. It has subsequently been extended to a wide range of stochastic processes with resetting \cite{Evans20}. One of the novel features of partially absorbing surfaces with non-constant reactivities is that one has to maintain a memory of the current boundary local time under reset in order to correctly account for the statistics of absorption. That is, the absorption probability depends on the number of encounters between particle and boundary, which has to be preserved during resetting. This means that the resetting protocol is not given by a renewal process \cite{Bressloff22b}. In order to simplify the analysis, we  introduced a modified resetting rule in which both the position of the particle and the local time were reset, which is governed by a renewal process.We also showed that the effects of a partially absorbing surface on the mean first passage time (MFPT) for total absorption differs significantly if local time resetting is included. That is, the MFPT for a totally absorbing surface is increased by a multiplicative factor when the local time is reset, whereas the MFPT is increased additively when only particle position is reset.

 \begin{figure}[b!]
  \raggedleft
  \includegraphics[width=10cm]{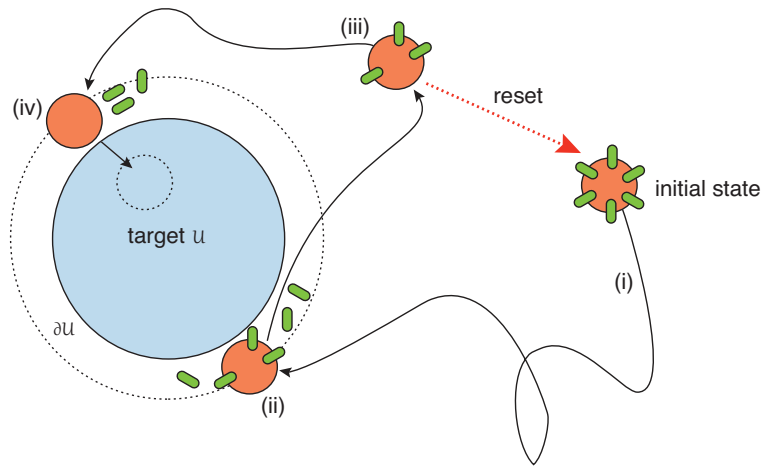}
  \caption{Hypothetical mechanism for threshold surface absorption. (i) A particle carrying a set of resources, say, diffuses in a domain containing a single target $\calU$. (For illustrative purposes we show a discrete set of resources.) (ii) Whenever the particle is in a neighborhood of the target surface $\partial \calU$ it can unload its resources at some specified rate. The total amount delivered is taken to be a monotonically increasing function of the boundary local time $\ell_t$. (iii) The particle continues to diffuse in the bulk until (iv) all of its resources have been unloaded, after which it is immediately absorbed. Prior to absorption, the particle can reset to its initial state at some fixed rate $r$.}
  \label{fig1}
\end{figure}

In this paper we further explore the effects of local time resetting based on a so-called threshold surface reaction in a domain containing a single target $\calU $. One possible mechanism for such a reaction is shown in Fig. \ref{fig1}. Suppose that a diffusing particle 
carries a packet of $h$ resources, which it can unload at some fixed rate whenever it is within a neighborhood of $\partial \calU$ see Fig. \ref{fig1}. Let $Z_t$ denote the amount of resources remaining at time $t$, such that absorption occurs as soon as all of the resources have been unloaded ($Z_t=0$). In addition, we take $Z_t$ to decrease monotonically with the boundary local time $\ell_t$ at the surface $\partial \calU$, that is, $Z_t=Z(\ell_t)$ with $Z'(\ell)<0$ for all $\ell \geq 0$ and $Z(0)=h$. (We assume that $h$ is sufficiently large so that $Z_t=Z(\ell_t)$ can be treated as a continuous function.)  It follows that absorption occurs as soon as the local time reaches the threshold $\ell_h=Z^{-1}(0)$, whereas the surface is totally reflecting when $0\leq \ell_t <\ell_h$. That is, the corresponding stopping local time distribution is a Heaviside function, $\Psi(\ell)=H(\ell_h-\ell)$. We can now define a resetting protocol in which the particle position is reset to its initial value at a Poisson rate $r$, $\X_t\rightarrow \x_0$, and the particle is immediately resupplied with its full complement of cargo, that is, $Z_t\rightarrow h$. This is mathematically equivalent to resetting the local time according to $\ell_t\rightarrow 0$. (If the particle load is not replenished then only the position is reset.)

 The structure of the paper is as follows. In section 2 we introduce the threshold surface reaction for diffusion
in a bounded domain $\Omega \subseteq \R^d$ containing a single target $\calU\subset \Omega$. The exterior boundary $\partial \Omega$ is taken to be totally reflecting, whereas the interior boundary $\partial \calU$ is a reactive surface that modifies the internal state of the particle as outlined above. We first consider diffusion without resetting and show how to formulate the threshold surface reaction in terms of the generalized propagator for reflected Brownian motion. We then show how to incorporate stochastic resetting into the propagator BVP with and without local time resetting, and determine the corresponding survival probabilities and MFPTs. We illustrate the theory using the simple examples of diffusion on the half-line (section 3) and a spherical target in a spherical domain (section 4).
  
  \section{Threshold surface reactions and the generalized propagator}

\subsection{Threshold surface reactions without resetting}

\begin{figure}[t!]
  \raggedleft
  \includegraphics[width=10cm]{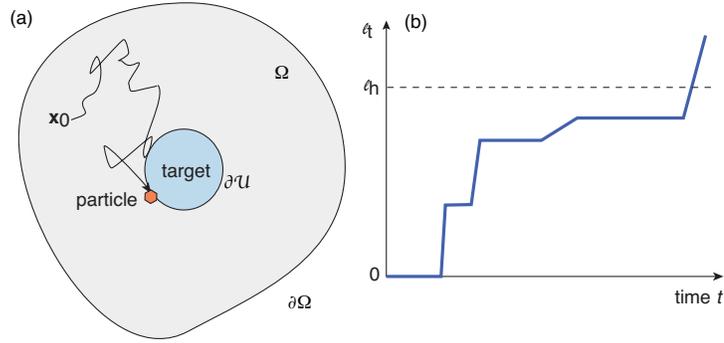}
  \caption{A single reactive target $\calU$ in the interior of a bounded domain $\Omega\subset \R^d$. The exterior boundary $\partial \Omega$ is totally reflecting. (a) A particle diffusing in $\Omega \backslash \calU$ interacts with the reactive surface $\partial \calU$ by modifying an internal state $Z_t$ with $Z_0=h$. The latter decreases monotonically with the time spent at the boundary, as specified by the boundary local time $\ell_t$, that is, $Z_t=Z(\ell_t)$ with $Z(0)=h$ and $Z'(\ell)<0$ for all $\ell$. If $0\leq Z_t<h$ then the surface is totally reflecting, whereas the particle is absorbed by the surface as soon as $Z_t$ reaches zero. (b) Schematic plot of the local time $\ell_t$. Absorption occurs as soon as $\ell_t$ crosses the threshold $\ell_h=Z^{-1}(0)$.}
  \label{fig2}
\end{figure}

Consider a particle diffusing in a bounded domain $\Omega$ containing an interior target $\calU$ with a reactive boundary $\partial \calU$, see Fig. \ref{fig2}(a). 
For the moment, suppose that both the exterior boundary $\partial \Omega$ and the interior surface $\partial \calU$ are totally reflecting. Let $\X_t \in \Omega\backslash \calU$ represent the position of the particle at time $t$. The boundary local time for the totally reflecting surface $\partial \calU$ is defined according to \cite{Levy39,McKean75,Majumdar05,Grebenkov19a}
\begin{equation}
\label{loc}
\ell_t=\lim_{\delta\rightarrow 0} \frac{D}{\delta} \int_0^tH(\delta-\mbox{dist}(\X_{\tau},\partial \calU))d\tau,
\end{equation}
where $H$ is the Heaviside function. Note that $\ell_t$ has units of length due to the additional factor of $D$, but can still be viewed as the amount of time that the particle spends in an infinitesimal neighborhood of the surface $\partial \calU$. It is clear from definition (\ref{loc}) that $\ell_t$ is a non-decreasing stochastic process, which remains at zero until the first encounter with the boundary. One well known property of reflected Brownian motion is that when a particle hits a smooth surface, it returns to the surface an infinite number of times within an infinitely short time interval. Although each of these returns generates an infinitesimal increase in the boundary local time, the net effect of multiple returns is a measurable change in $\ell_t$. In a real physical system, there is a natural surface boundary layer of width $\delta$ that is determined by short-range atomic interactions. One can then approximate the boundary local time by the residence time of the particle in the boundary layer. An analogous regularization occurs in numerical simulations due to spatial discretization. As highlighted in Ref. \cite{Grebenkov20}, the boundary local time is a more universal quantity since it is independent of the boundary layer width, and is thus easier to deal with mathematically.

Let $P_0(\x,\ell,t|\x_0)$ denote the joint probability density or propagator for the pair $(\X_t,\ell_t)$. (The subscript on $P_0$ indicates that there is no resetting.) The propagator satisfies a BVP that can be derived using integral representations \cite{Grebenkov20} or path-integrals \cite{Bressloff22a}. Here we briefly sketch the latter derivation. First, note that
 \begin{align}
 \label{A1}
& P_0(\x,\ell,t|\x_0)=\bigg \langle \delta\left (\ell -\ell_t \right )\bigg \rangle_{\X_0=\x_0}^{\X_t=\x} ,
 \end{align}
 where expectation is taken with respect to all random paths realized by $\X_{\tau}$ between $\X_0=\x_0$ and $\X_t=\x$. 
  Using a Fourier representation of the Dirac delta function, equation (\ref{A1}) can be rewritten as
 \begin{align}
 P_0(\x,\ell,t|\x_0)=\int_{-\infty}^{\infty} \e^{i\omega \ell}{\mathcal G}(\x,\omega,t|\x_0)\frac{d\omega}{2\pi},
 \end{align}
 where $ P_0(\x,u,t|\x_0)=0$ for $u<0$ and
 \begin{align}
 {\mathcal G}(\x,\omega,t|\x_0)=\bigg\langle \exp \left ( -i\omega \ell_t\right )\bigg \rangle_{\X_0=\x_0}^{\X_t=\x}.
 \end{align}
 We now note that ${\mathcal G}$ is the characteristic functional of $\ell_t$, whose path-integral representation can be used to derive the following equation:
\begin{align}
\label{calG}
 \frac{\partial \calG(\x,\omega,t|\x_0)}{\partial t}&=D\nabla^2 \calG(\x,\omega,t|\x_0) -i\omega  F(\x) \calG(\x,\omega,t|\x_0) ,
\end{align}
where
\begin{align}
F(\x)=\lim_{\delta\rightarrow 0} \frac{D}{\delta}H(\delta-\mbox{dist}(\x,\partial \calU))= D\int_{\partial \calU}\delta(\x-\x')d\x'.
\end{align}
Multiplying equation (\ref{calG}) by $\e^{\i\omega \ell}$, integrating with respect to $\omega$ and using the identity
\[\frac{\partial }{\partial \ell} P_0(\x,\ell,t|\x_0)H(\ell) =\int_{-\infty}^{\infty} i\omega  \e^{i\omega \ell}{\mathcal G}(\x,\omega,t|\x_0)\frac{d\omega}{2\pi},\]
 we obtain the equation
\begin{align}
& \frac{\partial P_0(\x,\ell,t|\x_0)}{\partial t}=D\nabla^2 P_0(\x,\ell,t|\x_0)-F(\x) \frac{\partial P_0}{\partial \ell}(\x,\ell,t|\x_0) \- \delta(\ell)F(\x)P_0(\x,0,t|\x_0) \nonumber \\
\label{calP}
 &=D\nabla^2 P_0(\x,\ell,t|\x_0) \\
 &\quad -D\int_{\partial \calU}\left (\frac{\partial P_0}{\partial \ell}(\x',\ell,t|\x_0) +\delta(\ell)P_0(\x,0,t|\x_0) \right )\delta(\x-\x')d\x',\  \x\in \R^d. \nonumber
\end{align}
Equation (\ref{calP})  is equivalent to the BVP
\begin{subequations}
\begin{align}
 &\frac{\partial P_0(\x,\ell,t|\x_0)}{\partial t}=D\nabla^2 P_0(\x,\ell,t|\x_0),\ \x \in \Omega\backslash \calU, \ \nabla P_0(\x,\ell,t|\x_0) \cdot \n =0,\,\x\in \partial \Omega,\nonumber \\
 & \label{Ploc1}\\
 &-D\nabla P_0(\x,\ell,t|\x_0) \cdot \n= D P_0(\x,\ell=0,t|\x_0) \ \delta(\ell)  +D\frac{\partial}{\partial \ell} P_0(\x,\ell,t|\x_0) ,\,  \x\in \partial \calU.\nonumber \\
\label{Ploc2}
\end{align}
Finally, using the equivalence of a Robin boundary condition and an exponential stopping local time distribution, it can be shown that \cite{Bressloff22a}
\begin{equation}
\label{Ploc3}
P_0(\x,\ell=0,t|\x_0)=-\nabla p_{0,\infty}(\x,t|\x_0)\cdot \n \mbox{ for } \x\in \partial \calU, 
\end{equation}
\end{subequations}
where $p_{0,\infty}$ is the probability density in the case of a totally absorbing target: 
\begin{subequations}
\begin{align}
 	&\frac{\partial p_{0,\infty}(\x,t|\x_0)}{\partial t} = D\nabla^2 p_{0,\infty}(\x,t|\x_0), \, \x\in \Omega\backslash \calU,\  \nabla p_{0,\infty}(\x,t|\x_0) \cdot \n=0 ,\,\x\in\partial \Omega,\nonumber \\
 &\label{pinf}
 \\
 &p_{0,\infty}(\x,t|\x_0)=0,\  \x\in \partial \calU,\ p_{0,\infty}(\x,0|\x_0)=\delta(\x-\x_0).
	\end{align}
	\end{subequations}
Equations (\ref{Ploc1})--(\ref{Ploc3}) are identical to the BVP derived in Ref. \cite{Grebenkov20} using a different method.

For future reference, it is convenient to Laplace transform the propagator BVP, which then takes the form
\begin{subequations}
\begin{align}
\label{Ploc1LT}
 & D\nabla^2 \PP_0(\x,\ell,s|\x_0)-s\PP_0(\x,\ell,s|\x_0)=-\delta(\ell)\delta(\x-\x_0),\ \x \in \Omega\backslash \calU,\nonumber \\
 &\\
\label{Ploc2LT} &-\nabla \PP_0(\x,\ell,s|\x_0) \cdot \n=0,\  \x\in \partial \Omega,\\
\label{Ploc3LT} &-\nabla \PP_0(\x,\ell,s|\x_0) \cdot \n=  \PP_0(\x,\ell=0,s|\x_0) \ \delta(\ell)  +\frac{\partial}{\partial \ell} \PP_0(\x,\ell,s|\x_0),\, \x\in \partial \calU,\\
\label{Ploc4LT}
 & \PP(\x,\ell=0,s|\x_0)=-\nabla \p_{0,\infty}(\x,s|\x_0)\cdot \n ,\ \x\in \partial \calU.
\end{align}
\end{subequations}
Furthermore, Laplace transforming equations (\ref{pinf}) shows that $\p_{0,\infty}=G(\x,s|\x_0)$ where $G$ is a modified Helmholtz Green's function:
\begin{align}
\label{GGa}
	&D\nabla^2 G(\x,s|\x_0)-sG(\x,s|\x_0)=-\delta(\x-\x_0), \, \x\in \Omega\backslash \calU,\nonumber \\
 &\nabla G(\x,s|\x_0)\cdot \n=0,\, \x\in \partial \Omega,\quad  G(\x,s|\x_0)=0,\  \x\in \partial \calU.
	\end{align} 

Now suppose that the particle has an internal state variable $Z_t$ that decreases monotonically with the local time $\ell_t$, that is, $Z_t=Z(\ell_t)$ with $Z(0)=Z_0=h$ and $Z'(\ell)<0$ for all $\ell \geq 0$. If $0\leq Z_t < h$, then the surface $\partial \calU$ remains totally reflecting. In other words, surface reactions modify the internal state of the particle without absorbing it. The particle is only absorbed by the surface when $Z_t$ reaches zero. We will refer to this form of absorption scheme as a threshold surface reaction.
It follows that the FPT ${\mathcal T}$ for particle absorption is, see Fig. \ref{fig2}(b),
\begin{equation}
\label{Tell0}
{\mathcal T}=\inf\{t>0:\ \ell_t >\ell_h \equiv Z^{-1}(0)>0\}.
\end{equation}
In addition, the marginal probability density for particle position is
\begin{equation}
 \label{Boo}
p_0(\x,t|\x_0)=\int_0^{\ell_h}P_0(\x,\ell,t|\x_0)d\ell.
\end{equation}
We thus have a particular example of the general class of diffusion-mediated surface reactions introduced in Ref. \cite{Grebenkov20}. That is, we can write
 \begin{equation}
  p_0(\x,t|\x_0)=\int_0^{\infty} \Psi(\ell)P_0(\x,\ell,t|\x_0)d\ell,
  \end{equation}
  with $\Psi(\ell)=H(\ell_h-\ell)$. (One can also interpret $p_0(\x,t|\x_0)$ as a survival probability density given the additional absorbing boundary condition $P_0(\x,\ell_h,t|\x_0) =0$.) As shown in Ref. \cite{Grebenkov20}, taking $\Psi(\ell) =\e^{-\gamma \ell}$ with $\gamma =\kappa_0/D$ recovers the classical Robin boundary condition  for a partially reflecting surface,
  \begin{equation}
  D\nabla p_0(\x,t|\x_0)\cdot \n +\kappa_0 p_0(\x,t|\x_0)=0,\quad \x \in \partial \calU,
  \end{equation}
where $\n$ is the unit normal at a point on $\partial \calU$ that is directed towards the interior of $\calU$. Other choices of $\Psi(\ell)$ allow for a partially absorbing surface whose reactivity depends on the accumulation time, that is, $\kappa_0\rightarrow \kappa(\ell)$.

In the case of a bounded domain $\Omega$, the local time $\ell_t$ crosses the threshold $\ell_h$ with probability one as $t\rightarrow \infty$. This means that
\begin{equation}
\lim_{t\rightarrow \infty} p_0(\x,t|\x_0)=0.
\end{equation}
The corresponding survival probability
\begin{equation}
\label{Qoo}
 Q_0(\x_0,t)\equiv \int_{\Omega\backslash \calU}p_0(\x,t|\x_0)d\x=\int_0^{\ell_h}\left [\int_{\Omega\backslash \calU}P_0(\x,\ell,t|\x_0)d\x\right ]d\ell
\end{equation}
satisfies  
\begin{align}
\frac{d Q_0(\x_0,t)}{dt}&=-D\int_{\partial  \calU}P_0(\x,\ell_h,t|\x_0)d\x\equiv -J_0(\x_0,t),
\end{align}
where $J_0(\x_0,t)$ is the probability flux into $\partial \calU$ at time $t$ in the absence of resetting. Laplace transforming this equation with $Q_0(\x_0,0)=1$ yields
\begin{equation}
s\Q_0(\x_0,s)-1=-\widetilde{J}_0(\x_0,s).
\end{equation}
Since $J_0(\x_0,t)$ is the FPT density for ${\mathcal T}$, the MFPT without resetting (if it exists) is given by
\begin{equation}
\label{T0}
T_0(\x_0)=\int_0^{\infty}tJ_0(\x_0,t)dt=-\left . \frac{\partial }{\partial s}\widetilde{J}_0(\x_0,s)\right |_{s=0}=\Q_0(\x_0,0).
\end{equation}

\subsection{Threshold surface reactions with position and local time resetting} 
\begin{figure}[t!]
  \raggedleft
  \includegraphics[width=10cm]{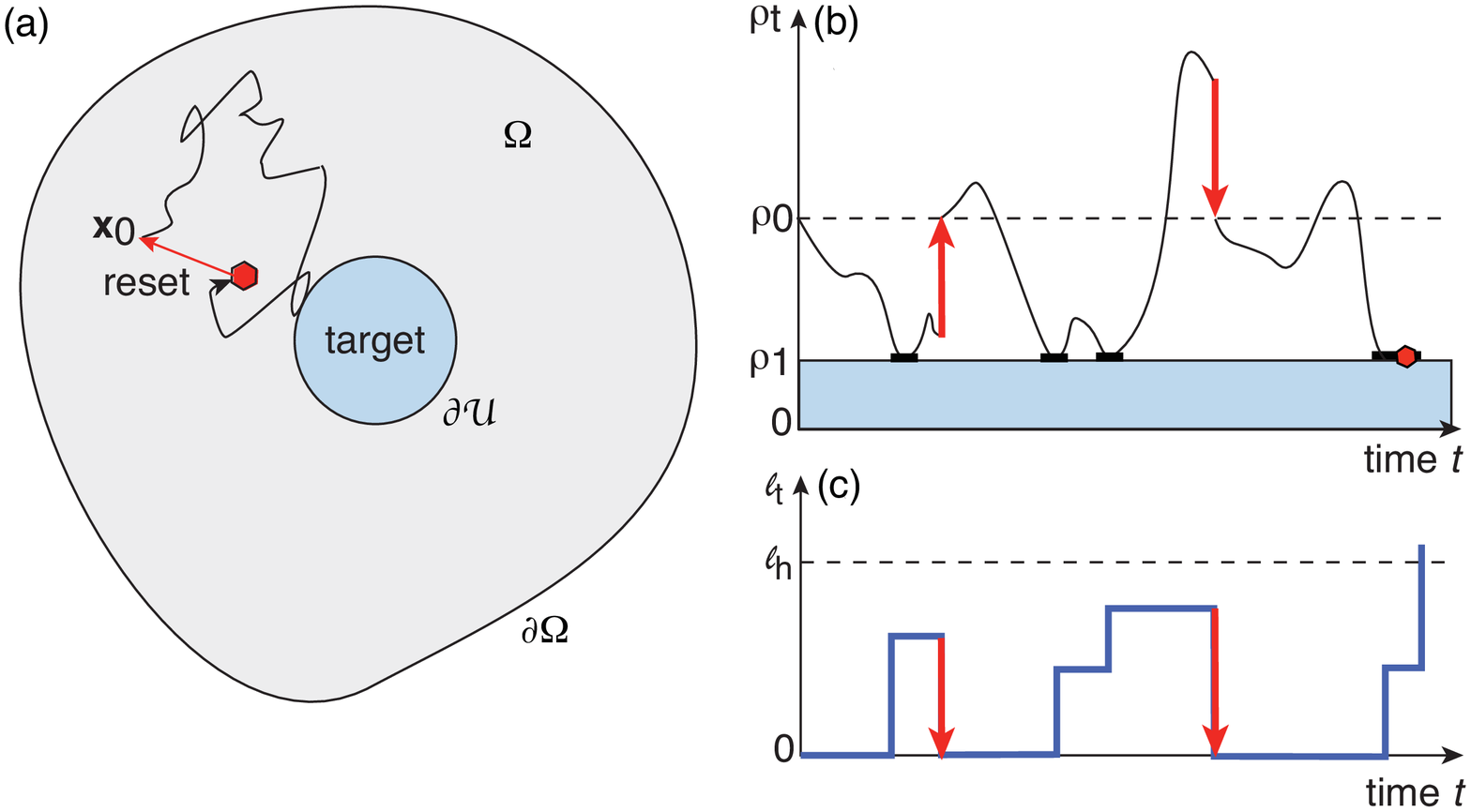}
  \caption{(a) Same as Fig. \ref{fig1}(a) except that now the particle can randomly reset its position $\X_t$ and internal state $Z_t$ at a rate $r$. (b) Schematic illustration of the variation in the distance $\rho_t$ from the center of a spherical target of radius $\rho_1$ with $|\x_0|=\rho_0>\rho_1$ and the resetting rule $\rho_t\rightarrow \rho_0$. (c) Corresponding time course of the local time $\ell_t$ under the resetting rule $\ell_t\rightarrow 0$.}
  \label{fig3}
\end{figure}

Prior to reaching the threshold for absorption, the surface is totally reflecting. This allows the introduction of a stochastic resetting protocol that restarts both the particle's position $\X_t$ and its internal state $Z_t$ irrespective of the current value of $\ell_t$, provided that $\ell_t < \ell_h$. Suppose that the final reset before absorption occurs at a time $\tau$. Then $Z_{\tau^+}=h$ and $Z_t =Z(\ell_t-\ell_{\tau})$ for $t>\tau$.
It follows that absorption will occur at the time ${\mathcal T}=\inf\{t>\tau:\ \ell_t >\ell_{\tau}+\ell_h\}$. Since this holds for all $\ell_{\tau}$, $\ell_{\tau}< \ell_h$, we can simply reset the local time according to $\ell_{\tau^+}=0$. Therefore, as a natural generalization of previous studies of diffusion processes with resetting \cite{Evans11a,Evans11b,Evans14}, we assume that at random times determined by a Poisson process with rate $r$ the pair $(\X_t,\ell_t)$ instantaneously resets to the state $(\x_0,0)$, after which diffusion is immediately resumed. The resetting scheme is illustrated in Fig. \ref{fig3}. It also follows that the resetting protocol is given by a renewal process with respect to the full state space $(\X_t,\ell_t)$.

Let $P_r(\x,\ell,t|\x_0)$ denote the propagator with resetting. Incorporating resetting into the BVP given by equations (\ref{Ploc1})--(\ref{Ploc3}), we have
\begin{subequations}
\begin{align}
 &\frac{\partial P_r(\x,\ell,t|\x_0)}{\partial t}=D\nabla^2 P_r(\x,\ell,t|\x_0)-rP_r(\x,\ell,t|\x_0)\nonumber \\
 &\hspace{3cm} +r\delta(\x-\x_0)\delta(\ell),\ \x \in \Omega\backslash \calU, \label{rPloc1} \\
 &-D\nabla P_r(\x,\ell,t|\x_0) \cdot \n= D P_r(\x,\ell=0,t|\x_0) \ \delta(\ell)  +D\frac{\partial}{\partial \ell} P_r(\x,\ell,t|\x_0) ,\,  \x\in \partial \calU,\nonumber \\
 \label{rPloc3}
\end{align}
\end{subequations}
The unknown $P_r(\x,\ell=0,t|\x_0) $ for $\x \in \partial \calU$ is determined by noting that the
stochastic resetting process is memoryless, and hence the propagator satisfies a first renewal equation of the form
\begin{equation}
 P_r(\x,\ell,t|\x_0)= \e^{-rt}P_0(\x,\ell,t|\x_0)+r\int_0^t\e^{-r \tau}P_r(\x,\ell,t-\tau|\x_0)d\tau .
\end{equation}
The first term on the right-hand side represents all trajectories that do not undergo any resettings, which occurs with probability $\e^{-rt}$. The second term represents the complementary set of trajectories that reset at least once with the first reset occurring at time $\tau$. Laplace transforming the renewal equation and rearranging shows that
\begin{equation}
\label{con}
\PP_r(\x,\ell,s|\x_0)=\left (1+\frac{r}{s}\right )\PP_0(\x,\ell,r+s|\x_0).
\end{equation}
Since $\PP_0(\x,\ell=0,s|\x_0)=-\nabla p_{0,\infty}(\x,s|\x_0) \cdot \n$ for $\x\in \calU$, it follows that
\begin{equation}
\PP_0(\x,\ell=0,s|\x_0)=-\left (1+\frac{r}{s}\right )\nabla p_{\infty}(\x,s|\x_0) \cdot \n,\quad \x \in \partial \calU.
\end{equation}
Multiplying both sides of equation (\ref{con}) by $s$ and taking the limit $s\rightarrow 0$, then establishes that there exists a non-equilibrium stationary state (NESS) $P_r^*(\x,\ell|\x_0)$:
\begin{equation}
\label{con0}
 P_r^*(\x,\ell|\x_0)=\lim_{t\rightarrow \infty}P_r(\x,\ell,t|\x_0)=\lim_{s\rightarrow 0}s \PP_r(\x,\ell,s|\x_0)=r\PP_0(\x,\ell,r|\x_0).
\end{equation}

The calculation of the associated marginal probability density with resetting is more involved than the case of no resetting. That is, one cannot simply take
$p_r(\x,t|\x_0)=\int_0^{\ell_h}P_r(\x,\ell,t|\x_0)d\ell$, since the set of paths $\{(\X_{\tau},\ell_{\tau}), \tau \in [0,t]\}$ that contribute to $P_r(\x,\ell,t|\x_0)$ include those that cross the threshold $\ell_h$ prior to resetting. In other words, the effective local time $\ell_t$ is no longer a monotonically increasing function of time. Therefore, we will proceed by partitioning the set of contributing paths according to the number of resettings, and explicitly exclude any paths that cross the threshold. Let ${\mathcal I}_t$ denote the number of resettings in the interval $[0,t]$ and let ${\mathcal T}=\inf\{t, \ell_t =\ell_h\}$. Then
\begin{align}
 p_r(\x,t|\x_0)d\x&=\e^{-rt}\P[\X_t \in [\x,\x+d\x]|\X_0=\x_0,\, {\mathcal T}>t,\, {\mathcal I}_t=0]\\
 &\quad +r\e^{-rt}\P[\X_t \in [\x,\x+d\x]|\X_0=\x_0,\, {\mathcal T}>t,\, {\mathcal I}_t=1]\nonumber \\
 &\quad +r^2\e^{-rt}\P[\X_t \in [\x,\x+d\x]|\X_0=\x_0,\, {\mathcal T}>t,\, {\mathcal I}_t=2]+\ldots \nonumber 
\end{align}
That is,
\begin{align}
 p_r(\x,t|\x_0)&=\e^{-rt} p_0(\x,t|\x_0)+r\e^{-rt} \int_0^t p_0(\x,\tau|\x_0)Q_0(\x_0,t-\tau)d\tau\\
 &\quad +r\e^{-rt} \int_0^t \int_0^{t-\tau}p_0(\x,\tau|x_0)Q_0(\x_0,t-\tau)Q_0(\x_0,t-\tau-\tau')d\tau'd\tau+\ldots \nonumber
\end{align}
where $Q_0$ is the survival probability without resetting, see equation (\ref{Qoo}). Laplace transforming the above equation and using the convolution theorem shows that
\begin{align}
 \p_r(\x,s|\x_0)&=\p_0(\x,r+s|\x_0)+r\p_0(\x,r+s|\x_0)\Q_0(x_0,r+s)\nonumber \\
&\quad +r^2\p_0(\x,r+s|\x_0)\Q_0(x_0,r+s)^2+\ldots
\end{align}
Summing the geometric series thus yields the result
\begin{equation}
\label{prQ}
\p_r(\x,s|\x_0)=\frac{\p_0(\x,r+s|\x_0)}{1-r\Q_0(\x_0,r+s)}.
\end{equation}
Moreover, integrating both sides with respect to $\x\in \Omega \backslash \calU$ implies that
\begin{equation}
\label{rQ}
\Q_r(\x_0,s)=\frac{\Q_0(\x_0,r+s)}{1-r\Q_0(\x_0,r+s)}.
\end{equation}
It immediately follows from the analog of equation (\ref{T0}) that the MFPT for absorption with resetting is
\begin{equation}
\label{Tr}
T_r(\x_0)=\Q_r(\x_0,0)=\frac{\Q_0(\x_0,r)}{1-r\Q_0(\x_0,r)}.
\end{equation}
This establishes that the MFPT can be obtained by explicitly solving the BVP for the generalized propagator without resetting,

\subsection{Threshold surface reactions with position resetting}

In a previous paper we considered diffusion-mediated surface reactions in which only the position of the particle resets \cite{Bressloff22b}. The propagator BVP becomes
\begin{subequations}
\begin{align}
\label{xrPloc1}
 &\frac{\partial P_r(\x,\ell,t|\x_0)}{\partial t}=D\nabla^2 P_r(\x,\ell,t|\x_0)-rP_r(\x,\ell,t|\x_0)\\
 &\hspace{3cm} +rQ_r(\x_0,\ell,t)\delta(\x-\x_0),\ \x \in \R^d\backslash \calU,\nonumber \\
\label{xrPloc2} &-D\nabla P_r(\x,\ell,t|\x_0) \cdot \n= D \delta(\ell)  P_r(\x,\ell=0,t|\x_0) +D\frac{\partial}{\partial \ell} P_r(\x,\ell,t|\x_0),\  \x\in \partial \calU,\\
\label{xrPloc3}
 & P_r(\x,\ell=0,t|\x_0)=-\nabla p_{r,\infty}(\x,t|\x_0)\cdot \n ,\ \x\in \partial \calU, 
\end{align}
\end{subequations}
where $P_r(\x,\ell,0|\x_0)=\delta(\x-\x_0)\delta(\ell)$ and $p_{r,\infty}$ is now the probability density in the case of a totally absorbing target with resetting: 
\begin{subequations} 
\begin{align}
\label{xpinf1}
 	&\frac{\partial p_{r,\infty}(\x,t|\x_0)}{\partial t} = D\nabla^2 p_{r,\infty}(\x,t|\x_0)-r p_{r,\infty}(\x,t|\x_0)\\
 & \hspace{3cm} \quad +rQ_{r,\infty}(\x_0,t) \delta(\x-\x_0), \, \x\in \R^d\backslash \calU,\nonumber \\
 &p_{r,\infty}(\x,t|\x_0)=0,\  \x\in \partial \calU,\ p_{r,\infty}(\x,0|\x_0)=\delta(\x-\x_0).
\label{xpinf2}
	\end{align}
	\end{subequations} 
	We have also introduced the marginal probabilities
	\begin{align}
 Q_r(\x_0,\ell,t)=\int_{\R^d\backslash \calU}P_r(\x,\ell,t|\x_0)d\x,\quad  Q_{r,\infty}(\x_0,t)=\int_{\Omega\backslash \calU}p_{r,\infty}(\x,t|\x_0)d\x.
	\end{align}
	
Since $\ell_t$ is now a monotonically increasing function of $t$, the corresponding survival probability can be obtained directly from $P_r$:
 \begin{equation}
  \label{oo}
   Q_r (\x_0,t)=\int_0^{\ell_h} Q_r(\x_0,\ell,t )d\ell=\int_0^{\ell_h}\left [\int_{\Omega\backslash \calU}P_r(\x,\ell,t|\x_0)d\x\right ]d\ell.
  \end{equation}
  However, the calculation of $P_r$ is non-trivial due to the fact that we do not have a renewal process. Therefore, we proceed along the lines of Ref. \cite{Bressloff22b} by considering the integral equation
\begin{align}
 P(\x,\ell,t|\x_0)&=\e^{-rt}P_0(\x,\ell,t|\x_0)\nonumber \\
 &\quad +r\int_0^{\ell} \left (\int_0^t \e^{-r\tau}P_0(\x,\ell-\ell',\tau|\x_0)Q_r(\x_0,\ell',t-\tau)d\tau \right )d\ell'.
\end{align}
The first term on the right-hand side represents all trajectories that do not reset in the interval $[0,t]$. The double integral represents the complementary set of trajectories that reset to $\x_0$ at least once. In particular, we assume that, prior to the last reset at time $t-\tau$, the particle spends a time $\ell'$ in a neighborhood of the boundary without being absorbed. This occurs with probability density $Q_r(\x_0,\ell',t-\tau)$. Over the time interval $[t-\tau,t]$ there are no more resettings and the local time increases by an additional amount $\ell-\ell'$ with associated probability density $P_0(\x,\ell-\ell',\tau|\x_0)$. Laplace transforming the integral equation using the convolution theorem shows that
\begin{align}
\label{okk}
 \PP(\x,\ell,s|\x_0)=\PP_0(\x,\ell,r+s|\x_0)+r\int_0^{\ell} \PP_0(\x,\ell-\ell',r+s|\x_0)Q_r(\x_0,\ell',s)d\ell' .
\end{align}
We now perform a second Laplace transform with respect to the local time $\ell$ by setting
\begin{subequations}
\begin{align}
 \calP(\x,z,s|\x_0)&=\int_0^{\infty}\e^{-z\ell}\PP(\x,\ell,s|\x_0)d\ell,\\
  \calQ_r(\x_0,z,s)&=\int_0^{\infty}\e^{-z\ell}\Q_r(\x_0,\ell,s)d\ell=\int_{\Omega\backslash \calU}\calP(\x,z,s|\x_0)d\x.
 \end{align}
 \end{subequations}
Multiplying both sides of equation (\ref{okk}) by $\e^{-z \ell}$ and applying the convolution theorem to the $\ell$-Laplace transform implies that
\begin{equation}
\label{calpspec}
\calP(\x,z,s|\x_0)=[1+r\calQ_r(\x_0,z,s)]\calP_0 (\x,z,r+s|\x_0).
\end{equation}
Integrating both sides of this equation with respect to $\x\in \Omega\backslash \calU$ gives
\begin{align}
\label{Qr0gen}
   \calQ_r(\x_0, z,s) =  [1+r\calQ_r(\x_0,z,s)]\calQ_0(\x_0,z,r+s),
\end{align} 
which can be rearranged so that
\begin{equation}
\label{Qgen}
\calQ_r(\x_0,z,s)=\frac{\calQ_0(\x_0,z,r+s)}{1-r\calQ_0(\x_0,z,r+s)}.
\end{equation}

The calculation of the survival probability of equation (\ref{oo}) thus proceeds in two steps. First, we solve the BVP for $\calP_0(\x,z,s|\x_0)$, which is obtained by Laplace transforming equations (\ref{Ploc1LT})--(\ref{Ploc4LT}) with respect to $\ell$:
\begin{subequations}
\begin{align}
\label{zPloc1LT}
 & D\nabla^2 \calP_0(\x,z,s|\x_0)-s\calP_0(\x,z,s|\x_0)=-\delta(\x-\x_0),\ \x \in \Omega\backslash \calU,  \\
\label{zPloc2LT} &-\nabla \calP_0(\x,z,s|\x_0) \cdot \n=0,\  \x\in \partial \Omega,\\
\label{zPloc3LT} &-\nabla \calP_0(\x,z,s|\x_0) \cdot \n=  z \calP_0(\x,z,s|\x_0),\, \x\in \partial \calU.
\end{align}
\end{subequations}
This is identical in form to the BVP for the probability density $\p_0(\x,s|\x_0) $ of a particle diffusing in a domain with a Robin boundary condition on the surface $\partial \calU$ with $z$ the constant rate of absorption. Alternatively, we simply take the $z$-Laplace transform of the solution $\PP_0(\x,\ell,s|\x_0)$ of the BVP (\ref{Ploc1LT})--(\ref{Ploc4LT}). Having obtianed $\calP_0(\x,z,s|\x_0)$, we integrate with respect to $\x\in \Omega\backslash \calU$ to generate $\calQ_0(\x_0,z,s)$. Second, substituting for $\calQ_0(\x_0,z,s)$ into equation (\ref{Qgen}), we determine $\Q_r(\x_0,\ell,s)$
 by inverting the Laplace transform with respect to $z$. 
The corresponding MFPT can then be calculated according to
\begin{align}
T_r(\x_0)&=\int_{0}^{\ell_h} \Q_r(\x_0,\ell,0)d\ell.
\label{Trl}
\end{align}

\setcounter{equation}{0}
\section{Semi-infinite interval}

As a simple illustration of the above theory, consider diffusion in the semi-infinite domain $[0,\infty)$, with a threshold absorbing boundary at $x=0$. The first step is to solve the BVP for $\PP_0(x,\ell, s|x_0)$, given by the 1D version of equations (\ref{Ploc1LT})--(\ref{Ploc4LT}):
  \begin{subequations}
\begin{align}
\label{rspha1D}
   &D\frac{\partial^2\PP_0(x,\ell, s|x_0)}{\partial x^2}  -s\PP_0(x,\ell, s|x_0) =-\delta(\ell)\delta(x-x_0) ,   0<x<\infty,\\
 \label{rsphb1D}
 &\frac{\partial }{\partial x}\PP_0(x,\ell,s|x_0) =  \PP_0(x,\ell=0,s|x_0) \ \delta(\ell)  +\frac{\partial}{\partial \ell} \PP_r0x,\ell,s|x_0),\  x=0,\\
  &  \PP_0(0,\ell=0,s|x_0)=\left . \frac{dG(x,s|x_0)}{dx}\right |_{x=0}.
 \label{rsphc1D}
\end{align}
\end{subequations}
Here $G$ is the modified Helmholtz Green's function in the case of a totally absorbing boundary condition at $x=0$:
\begin{align}
\label{Ga1D}
&D\frac{\partial^2G}{\partial x^2}  -sG  = -\delta(x - x_0), \ 0<x<\infty,\quad G(0,s|x_0)=0.
\end{align}
That is,
\begin{align}
\label{GG1D}
G(x, s|x_0) = \frac{1}{2\sqrt{sD} }\left [\e^{-\sqrt{s/D} |x-x_0|}-\e^{-\sqrt{s/D} |x+x_0|}\right ].
 \end{align}
The general solution of equations (\ref{rspha1D})--(\ref{rsphc1D}) is of the form
\begin{equation}
\label{1Dgir}
 \PP_0(x,\ell,s|x_0)=A(\ell,s)\e^{-\alpha x}+\delta(\ell) G(x,s|x_0), \quad  \alpha=\sqrt{\frac{s}{D}}.
\end{equation}
The first term on the right-hand side of equation (\ref{1Dgir}) is the solution to the homogeneous version of equation (\ref{rspha1D}).  

The unknown coefficient $A(\ell,s)$ is determined by imposing the boundary condition (\ref{rsphb1D}):
 \begin{align}
\label{1DCB2}
  &\frac{dA(\ell,s)}{d\ell}+\alpha A(\ell,s)=0\end{align}
Hence, $A(\ell,s)=A(0,s)\e^{-\alpha \ell}$ with $A(0,s)$ determined by setting $\ell=0$ and $x=0$ in equation (\ref{1Dgir}) and using (\ref{rsphc1D}):
We thus obtain the the following solution for the Laplace transformed propagator:
\begin{align}
\PP_0(x,\ell,s|x_0)&= \frac{1}{D} \e^{ -\alpha \ell} \e^{-\alpha (x+x_0)}+\delta(\ell)G(x, s|x_0) .
\end{align}
Integrating with respect to $x\in [0,\infty)$ yields
\begin{equation}
\label{1DQ}
 \Q_0(x_0,\ell,s)= \int_0^{\infty} \PP_0(x,\ell,s|x_0)dx=\frac{1}{\sqrt{sD}} \e^{ -\sqrt{s/D} (\ell+x_0)} +\delta(\ell)\int_0^{\infty}G(x, s|x_0)dx .
\end{equation}
In addition, integrating equation (\ref{Ga1D}) with respect to $x$ shows that
\begin{equation}
s\int_0^{\infty} G(x,s|x_0)dx=1-D\partial_xG(0,s|x_0)=1- \e^{-\sqrt{s/D}x_0}  .
\end{equation}
Combining these results, we find that the Laplace transformed survival probability for threshold surface absorption without resetting is
\begin{equation}
\Q_0(x_0,s)=\frac{1}{s}\left (1-\e^{-\alpha [x_0+\ell_h]}\right ).
\end{equation}

Now suppose that we include position and local time resetting. It immediately follows from equation (\ref{Tr}) that the MFPT for absorption is
\begin{equation}
\label{Tr1D}
T_r(x_0)=\frac{\Q_0(x_0,r)}{1-r\Q_0(x_0,r)}=\frac{1}{r}\left (\e^{\sqrt{r/D} [x_0+\ell_h]}-1\right ).
\end{equation}
This is identical to the expression derived for a totally absorbing boundary \cite{Evans11a,Evans11b} under the mapping $x_0\rightarrow x_0+\ell_h$. Hence, the threshold $\ell_h$ effectively shifts the starting position. Note that $T_r(x_0)\rightarrow 0$ for $r\rightarrow 0$ and $r\rightarrow \infty$, and there exists a unique minimum at the optimal resetting rate $r_{\rm opt}(x_0+\ell_h)$, where $r_{\rm opt}(x_0)$ is the corresponding optimum for a totally absorbing boundary.

\begin{figure}[b!]
  \raggedleft
  \includegraphics[width=8cm]{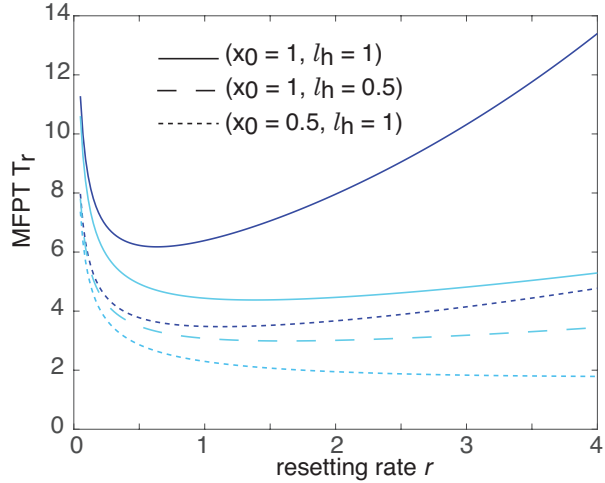}
  \caption{Plots of the MFPT $T_r$ as a function of the resetting rate $r$ for diffusion on the half-line with either position resetting (lighter curves) or position and local time resetting (darker curves). Various combinations of the initial position $x_0$ and local time threshold $\ell_h$ are considered. In the case of position resetting the MFPT only depends on the sum $x_0+\ell_h$. Time is in units of $1/r$, so that the units of length are fixed by taking $D=1$.}
  \label{fig4}
\end{figure}

\begin{figure}[t!]
  \raggedleft
  \includegraphics[width=8cm]{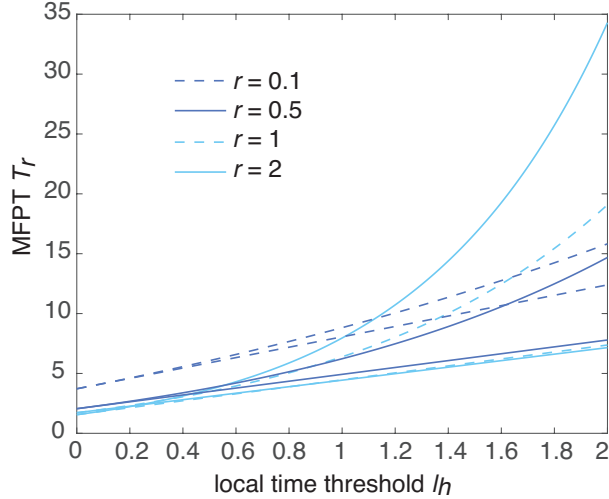}
  \caption{Plots of the MFPT $T_r$ as a function of the local time threshold $\ell_h$ for diffusion on the half-line and various resetting rates $r$. For each $r$, the upper curve is for position and local time resetting whereas the lower curve only includes position resetting. Other parameters are $x_0=D=1$.}
  \label{fig5}
  \end{figure}

On the other hand, if only the position resets, then the MFPT is given by
\begin{equation}
T_r(x_0)=\int_0^{\ell_h}{\mathcal L}^{-1}\left [\Phi(z)\right ](\ell) d\ell
\end{equation}
where ${\mathcal L}^{-1}$ is the inverse Laplace transform with respect to $z$,
\begin{align}
\Phi(z)&\equiv\frac{\calQ_0(x_0,z,r)}{1-r\calQ_0(x_0,z,r)} \end{align}
for fixed $r,x_0$, and
\begin{align}
 \calQ_0(x_0,z,r)&=\int_0^{\infty}\e^{-z\ell}\Q_0(x_0,\ell,s)d\ell\\
 &=\frac{1}{\sqrt{rD}}\frac{\e^{-\sqrt{r/D}x_0}}{z+\sqrt{r/D}}+\frac{1}{r}\left (1- \e^{-\sqrt{r/D}x_0}  \right ).
\end{align}
It follows that
\begin{align}
\Phi(z)&= \frac{\e^{-\sqrt{r/D}x_0}}{z\sqrt{rD}}+\frac{1}{r}\left (\e^{\sqrt{r/D} x_0}-1\right )
\end{align}
and
\begin{equation}
{\mathcal L}^{-1}\left [\Phi(z)\right ](\ell) =\frac{\e^{-\sqrt{r/D}x_0}}{\sqrt{rD}}+\frac{1}{r}\left (\e^{\sqrt{r/D} x_0}-1\right )\delta(\ell).
\end{equation}
Hence,
\begin{equation}
\label{Tr1Da}
T_r(x_0)=\frac{\ell_h }{\sqrt{rD}}\e^{\sqrt{r/D}x_0}+\frac{1}{r}\left (\e^{\sqrt{r/D} x_0}-1\right )
\end{equation}
Comparing equation (\ref{Tr1D}) with (\ref{Tr1Da}) shows that the effects of a partially absorbing surface on the mean first passage time (MFPT) for total absorption differs significantly if local time resetting is included. That is, the MFPT for a totally absorbing surface is increased by a multiplicative factor when the local time is reset, whereas the MFPT is increased additively when only particle position is reset. Example plots of $T_r$ as a function of the resetting rate $r$ are shown in Fig. \ref{fig4}. It can be seen that for given $x_0$ and $\ell_h$, the inclusion of local time resetting increases the MFPT as a function of $r$, and shifts the minimum of the resetting curve to the left. Hence, local time resetting reduces the optimal resetting rate $r_{\rm opt}$ at which $T_r$ is minimized. In Fig. \ref{fig5} we show corresponding plots of $T_r$ as a function of $\ell_h$ and fixed $r$. As expected, $T_r$ increases linearly with $\ell_h$ for position resetting (additive behavior), whereas it increases exponentially with $\ell_h$ when local time resetting is included (multiplicative behavior).

\setcounter{equation}{0}
\section{Spherical target in a spherical domain}

Let us now consider a spherical domain $\Omega =\{\x\in \R^d\,|\, 0\leq  |\x| <\rho_2\}$ and a spherical target of radius $\rho_1$ at the center of $\Omega$ with $\rho_1<\rho_2$:
\[\calU=\{\x\in \R^d \, |\, 0\leq  |\x|<\rho_1\},\quad \partial \calU=\{\x\in \R^d\, |\, |\x|=\rho_1\}.\]
(In the 1D case ($d=1$) we obtain the finite-interval version of the example considered in section 3, with a partially absorbing boundary at $x=\rho_1$ and a reflecting boundary at $x=\rho_2$.)
Following \cite{Redner01}, the initial position of the particle is randomly chosen from the surface of the sphere of radius $\rho_0$, $\rho_1<\rho_0<\rho_2$.  That is,
\begin{align}
    P_0({\bf x}, \ell,0|{\bf x}_0) = \Gamma_d\delta(\rho - \rho_0)\delta(\ell),\quad \Gamma_d=\frac{1}{\Omega_d \rho_0^{d - 1}} ,
\end{align}
where $\rho= \|{\bf x}\|$ and $\Omega_d$ is the surface area of a unit sphere in $\mathbb{R}^d$. This allows us to exploit spherical symmetry such that  $P_0=P_0(\rho,\ell,t|\rho_0)$. Introducing spherical polar coordinates, we can write equations (\ref{Ploc1LT})--(\ref{Ploc4LT}) as
  \begin{subequations}
\begin{align}
   &D\frac{\partial^2\PP_0}{\partial \rho^2} + D\frac{d - 1}{\rho}\frac{\partial \PP_0}{\partial \rho} -s\PP_0(\rho,\ell, s|\rho_0) =-\delta(\ell) \Gamma_d\delta(\rho-\rho_0) ,\   \rho_1<\rho<\rho_2,
 \label{rspha}
 \\
 \label{rsphb0}
  & \left . \frac{\partial }{\partial \rho}\PP_0(\rho,\ell,s|\rho_0)\right |_{\rho=\rho_2}=0,\\
 &\frac{\partial }{\partial \rho}\PP_0(\rho,\ell,s|\rho_0) =  \PP_0(\rho,\ell=0,s|\rho_0) \ \delta(\ell)  +\frac{\partial}{\partial \ell} \PP_0(\rho,\ell,s|\rho_0),\  \rho=\rho_1,
 \label{rsphb}\\
  & \PP_0(\rho_1,\ell=0,s|\rho_0)=\left . \frac{d}{d\rho}G(\rho,s|\rho_0) \right |_{\rho=\rho_1} .
 \label{rsphc}
\end{align}
\end{subequations}
Now
 $G$ is the modified Helmholtz Green's function satisfying
  \begin{subequations}
\begin{align}
\label{Ga}
    &D\frac{\partial^2G}{\partial \rho^2} + D\frac{d - 1}{\rho}\frac{\partial G}{\partial \rho} -sG  = -\Gamma_d \delta(\rho - \rho_0), \ \rho_1<\rho <\rho_2,\\ 
  & G(\rho_1,\alpha|\rho_0)=0,\quad \left .\frac{\partial }{\partial \rho}G(\rho,\alpha|\rho_0)\right |_{\rho=\rho_2}=0.
 \label{Gb}
\end{align}
\end{subequations}
The latter is given by \cite{Redner01}
\begin{align}
\label{GG}
  G(\rho, \alpha| \rho_0) = \frac{ (\rho\rho_0)^\nu }{D\Omega_d}\frac{C_{\nu}(\rho_{<},\rho_1;\alpha)D_{\nu}(\rho_{>},\rho_2;\alpha)}{D_{\nu}(\rho_1,\rho_2;\alpha)},
\end{align}
where $\rho_< = \min{(\rho, \rho_0)}$, $\rho_> = \max{(\rho, \rho_0)}$, and
\begin{subequations}
\begin{align}
 C_{\nu}(a,b;\alpha)&=I_{\nu}(\alpha a)K_{\nu}(\alpha b)-I_{\nu}(\alpha b)K_{\nu}(\alpha a),\\
  D_{\nu}(a,b;\alpha)&=I_{\nu}(\alpha a)K_{\nu-1}(\alpha b)+I_{\nu-1}(\alpha b)K_{\nu}(\alpha a).
\end{align}
\end{subequations}
Here $\nu=1-d/2$, $\alpha=\sqrt{s/D}$ and $I_{\nu},K_{\nu}$ are modified Bessel functions of the first and second kind, respectively.

The above BVP was previously solved in Ref. \cite{Bressloff22a} so we simply write down the final result:
\begin{align}
 \PP_0(\rho,\ell,s|\rho_0)&= \left [F_K(\rho,s)-F_I(\rho,s)\frac{F_K'(\rho_2,s)}{F_I'(\rho_2,s)}\right ]A_0(s)\e^{-\Lambda (s)\ell}+G(\rho,s|\rho_0)\delta(\ell),
\end{align}
where
\begin{equation}
F_I(\rho,s)=\rho^\nu I_\nu(\alpha \rho),\quad F_K(\rho,s)=\rho^\nu K_\nu(\alpha \rho).
\end{equation}
are solutions of the homogeneous diffusion equation in Laplace space,
\begin{align}
\label{GamR}
\Lambda(s)=-\frac{F_K'(\rho_1,s)F_I'(\rho_2,s)-F_I'(\rho_1,s)F_K'(\rho_2,s)}{F_K(\rho_1,s)F_I'(\rho_2,s)-F_I(\rho_1)F_K'(\rho_2,s)}.
\end{align}
and
\begin{equation}
A_0(s)= \left (F_K(\rho_1,s)-F_I(\rho_1,s)\frac{F_K'(\rho_2,s)}{F_I'(\rho_2,s)}\right )^{-1}\partial_{\rho}G(\rho_1,s|\rho_0).
\end{equation}
Integrating $\PP_0(\rho,\ell,s|\rho_0)$ over the annular region $\rho_1<\rho<\rho_2$ then gives
\begin{align}
 \Q_0(\rho_0,\ell,s)&=\Omega_d \int_{\rho_1}^{\rho_2} \rho^{d-1} \PP_0(\rho,\ell,s|\rho_0)d\rho\nonumber \\
 &= \left [\overline{F}_K(s)-\overline{F}_I(s) \frac{F_K'(\rho_2,s)}{F_I'(\rho_2,s)}\right ]A_0(s)\e^{-\Lambda (s)\ell}+\overline{G}(\rho_0,s )\delta(\ell),
\end{align}
with
\begin{equation}
\overline{f}\equiv \Omega_d \int_{\rho_1}^{\rho_2} \rho^{d-1}f(\rho)d\rho
\end{equation}
for any integrable function $f(\rho)$. In addition, integrating equation (\ref{Ga}) with respect to $\rho$ and using the divergence theorem shows that
\begin{equation}
s\overline{G}(\rho_0,s)=1-\Omega_d\rho_1^{d-1}D\partial_{\rho}G(\rho_1,s|\rho_0).
\end{equation}
Similarly, exploiting the fact that $F_{I,K}$ are homogeneous solutions of the spherically symmetric diffusion equation in Laplace space,
\begin{equation}
s\overline{F}_{I,K}(s)=-\Omega_d\rho_1^{d-1}DF_{I,K}'(\rho_1,s),
\end{equation}
which implies that
\begin{align}
\Q_0(\rho_0,\ell,s) &=\chi(\rho_0,s)\left ( \frac{\Lambda(s)\e^{-\Lambda (s)\ell}}{s}-\frac{\delta(\ell)}{s}\right )+\frac{\delta(\ell)}{s},
\end{align}
where
\begin{equation}
\chi(\rho_0,s)\equiv \Omega_d\rho_1^{d-1}D\partial_{\rho}G(\rho_1,s|\rho_0)=\left (\frac{\rho_0}{\rho_1}\right )^{\nu}\frac{D_{\nu}(\rho_0,\rho_2)}{D_{\nu}(\rho_1,\rho_2)}
\end{equation}
is the Laplace transformed flux into a totally absorbing surface.
Finally, integrating with respect to $\ell\in [0,\ell_h]$ yields the Laplace transformed survival probability
\begin{align}
 \Q_0(\rho_0,s) &=\frac{1}{s}\left (  1-\chi(\rho_0,s)\e^{-\Lambda (s)\ell_h}\right ).
\end{align}

\begin{figure}[b!]
  \raggedleft
  \includegraphics[width=8cm]{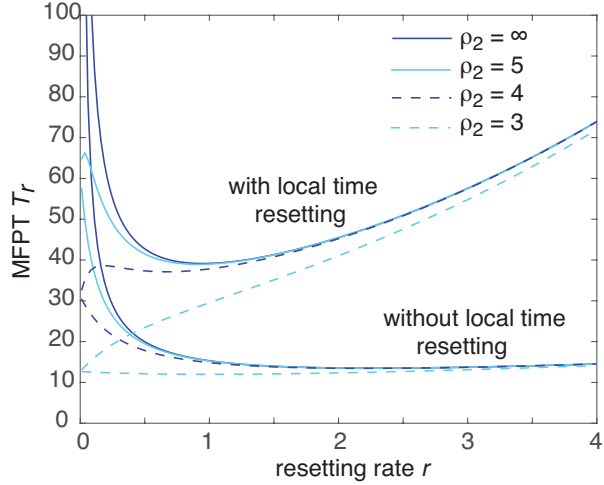}
  \caption{Diffusion in a 3D annular domain with a totally reflecting outer boundary of radius $\rho_2$ and a threshold absorbing inner boundary of radius $\rho_1$. Plots of the MFPT $T_r$ as a function of the resetting rate $r$ for various outer radii $\rho_2$ with either position resetting (lighter curves) or position and local time resetting (darker curves). Time is in units of $1/r$, so that the units of length are fixed by taking $D=1$. Other parameter values are $\ell_h=1$, $\rho_0=2$ and $\rho_1=1$.}
  \label{fig6}
\end{figure}

Equation (\ref{Tr}) implies that the MFPT for absorption by a spherical surface with position and local time resetting is
\begin{equation}
\label{Tr2D}
T_r(\rho_0)=\frac{\Q_0(\rho_0,r)}{1-r\Q_0(\rho_0,r)}=\frac{1}{r}\left (\frac{\e^{\Lambda (r)\ell_h}}{\chi(\rho_0,r)}-1\right ).
\end{equation}
In contrast, if only the position resets, then the MFPT is given by
\begin{equation}
T_r(\rho_0)=\int_0^{\ell_h}{\mathcal L}^{-1}\left [\Phi(z)\right ](\ell) d\ell
\end{equation}
where
\begin{align}
\Phi(z)&\equiv\frac{\calQ_0(\rho_0,z,r)}{1-r\calQ_0(\rho_0,z,r)} \end{align}
for fixed $r,x_0$, and
\begin{align}
 \calQ_0(\rho_0,z,r)&=\int_0^{\infty}\e^{-z\ell}\Q_0(\rho_0,\ell,s)d\ell\\
 &=\frac{\Lambda(r)}{r}\frac{\chi(\rho_0,r)}{z+\Lambda(r)}+\frac{1}{r}\left (1- \chi(\rho_0,r)  \right ).
\end{align}
It follows that
\begin{align}
\Phi(z)&= \frac{\Lambda(r)}{zr\chi(\rho_0,r)}+\frac{1}{r}\left (\frac{1}{\chi(\rho_0,r) }- 1 \right )
\end{align}
and
\begin{equation}
{\mathcal L}^{-1}\left [\Phi(z)\right ](\ell) =\frac{\Lambda(r)}{r\chi(\rho_0,r)}+\frac{1}{r}\left (\frac{1}{\chi(\rho_0,r) }- 1 \right )\delta(\ell).
\end{equation}
Hence,
\begin{equation}
\label{Tr2Da}
T_r(x_0)=\frac{\ell_h\Lambda(r)}{r\chi(\rho_0,r)}+\frac{1}{r}\left (\frac{1}{\chi(\rho_0,r) }- 1 \right ).
\end{equation}

\begin{figure}[t!]
  \raggedleft
  \includegraphics[width=8cm]{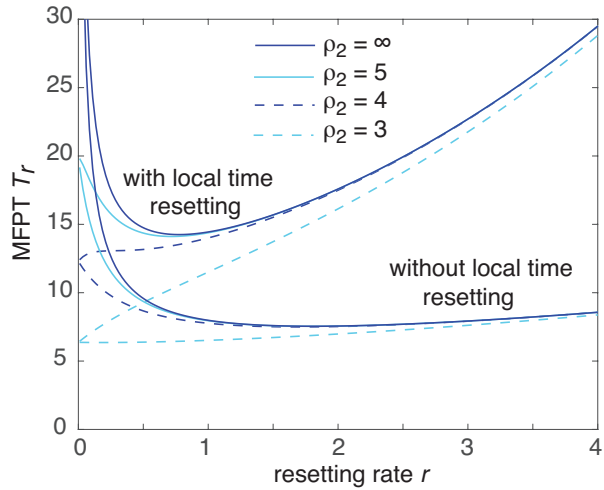}
  \caption{Same as Fig. \ref{fig6} for a 2D annular region.}
  \label{fig7}
  \end{figure}

\begin{figure}[t!]
  \raggedleft
  \includegraphics[width=8cm]{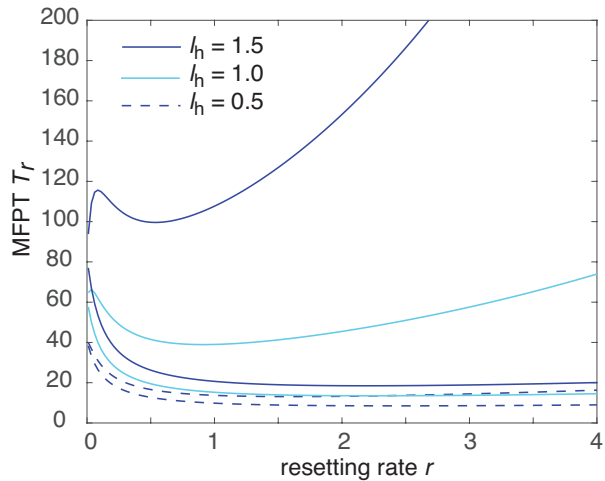}
  \caption{Same as Fig. \ref{fig6} for  various thresholds $\ell_h$ and $\rho_2=5$.}
  \label{fig8}
  \end{figure}

Example plots of $T_r$ as a function of the resetting rate $r$ for $d=3$ and $d=2$ are shown in Fig. \ref{fig6} and Fig. \ref{fig7}, respectively. Consistent with the results for diffusion on the half-line, the inclusion of resetting increases the MFPT and reduces the optimal resetting rate $r_{\rm opt}$. However, one major difference is that for $\rho_2 < \infty$, we have a bounded annular region, which means that the MFPT remains finite as $r\rightarrow 0$. As is well known in the case of diffusion in finite intervals \cite{Pal19}, $T_r$ may either be a monotonically increasing or unimodal function of $r$, depending on the size of the domain. This is clearly seen in Figs. \ref{fig6} and \ref{fig7}. As an additional consistency check, $\lim_{r\rightarrow 0}T_r$ is the same whether or not the local time is also reset. Finally, in Fig. \ref{fig8} we consider the effect of changing the local time threshold $\ell_h$. Again, as expected, when local time resetting is included, the MFPT curves are much more sensitive to variations in $\ell_h$.

\section{Discussion}  

In this paper we considered a hypothetical mechanism for threshold surface absorption, see Fig. \ref{fig1}, based on the idea that interactions with a target surface lead to the depletion of resources (or shrinkage) of a particle, such that once a critical level is reached the particle is immediately absorbed. This, in turn, motivated a novel resetting protocol in which both the position and boundary local time of the particle are reset. 

To what extent threshold surface absorption can be implemented, either naturally or artificially, remains to be seen. However, from a theoretical perspective, it widens the class of resetting protocols to include Brownian functionals. That is, although we focused on the boundary local time $\ell_t$ associated with the target surface $\partial \calU$, it is also possible to consider resetting of the corresponding occupation time $A_t$ in cases where the whole target domain $\calU$ is reactive rather than the boundary $\partial \calU$ \cite{Schumm21}. This means that the particle freely enters and exits $\calU$, and its internal state $Z_t$ is now a monotonically increasing function of $A_t$, which is a Brownian functional that specifies how long the particle spends within $\calU$ up to time $t$. More specifically, $A_t=\int_{0}^tI_{\calU}(\X_{\tau})d\tau $,
where $I_{\calU}(\x)$ denotes the indicator function of the set $\calU\subset \R^d$, that is, $I_{\calU}(\x)=1$ if $\x\in \calU$ and is zero otherwise. 
As we have recently shown elsewhere \cite{Bressloff22a}, one can derive a BVP for the associated propagator and incorporate partial absorption using a stopping occupation time distribution. Stochastic resetting can then be incorporated into the BVP along analogous lines to this paper.  The details will be presented elsewhere.  

Finally, as with other models of stochastic resetting \cite{Evans20}, the resetting processes considered in this paper are an idealization of more realistic active processes in which a particle returns to its initial position at some finite speed  \cite{Pal19a,Mendez19,Bodrova20,Pal20}, and there is a refractory period before the particle starts diffusing again \cite{Evans19a,Mendez19a}; the latter could represent the time needed to resupply the particle with cargo in Fig. \ref{fig1}. For simplicity, we assumed that resetting is instantaneous and ignored the effects of refractory periods. It would be interesting to explore the effects of delays in future work.

\end{document}